\documentclass[10pt, conference, compsocconf]{IEEEtran}
\ifCLASSINFOpdf
\else
\fi

\usepackage{graphicx}
\usepackage{url}
\usepackage{mdwlist} 
\usepackage{flushend}

\begin{document}
%
\title{A Factor Framework for Experimental Design for Performance Evaluation of Commercial Cloud Services}


\author{\IEEEauthorblockN{Zheng Li}
\IEEEauthorblockA{School of CS\\
NICTA and ANU\\
Canberra, Australia\\
Zheng.Li@nicta.com.au}
\and
\IEEEauthorblockN{Liam O'Brien}
\IEEEauthorblockA{Research School of CS\\
CECS, ANU\\
Canberra, Australia\\
liamob99@hotmail.com}
\and
\IEEEauthorblockN{He Zhang}
\IEEEauthorblockA{School of CSE\\
NICTA and UNSW\\
Sydney, Australia\\
He.Zhang@nicta.com.au}
\and
\IEEEauthorblockN{Rainbow Cai}
\IEEEauthorblockA{Division of Information\\
Information Services, ANU\\
Canberra, Australia\\
Rainbow.Cai@anu.edu.au}
}


%


\maketitle

\begin{abstract}
Given the diversity of commercial Cloud services, performance evaluations of candidate services would be crucial and beneficial for both service customers (e.g. cost-benefit analysis) and providers (e.g. direction of service improvement). Before an evaluation implementation, the selection of suitable factors (also called parameters or variables) plays a prerequisite role in designing evaluation experiments. However, there seems a lack of systematic approaches to factor selection for Cloud services performance evaluation. In other words, evaluators randomly and intuitively concerned experimental factors in most of the existing evaluation studies. Based on our previous taxonomy and modeling work, this paper proposes a factor framework for experimental design for performance evaluation of commercial Cloud services. This framework capsules the state-of-the-practice of performance evaluation factors that people currently take into account in the Cloud Computing domain, and in turn can help facilitate designing new experiments for evaluating Cloud services.

\end{abstract}


%
\IEEEpeerreviewmaketitle

\section{Introduction}
Along with the boom in Cloud Computing, an increasing number of commercial providers have started to offer public Cloud services \cite{Li_Yang_2010,Prodan_Ostermann_2009}. Since different commercial Cloud services may be supplied with different terminologies, definitions, and goals \cite{Prodan_Ostermann_2009}, performance evaluation of those services would be crucial and beneficial for both service customers (e.g. cost-benefit analysis) and providers (e.g. direction of improvement) \cite{Li_Yang_2010}. Before implementing performance evaluation, a proper set of experiments must be designed, while the relevant factors that may influence performance play a prerequisite role in designing evaluation experiments \cite{Jain_1991}. In general, one experiment should take into account more than one factor related to both the service to be evaluated and the workload. 

After exploring the existing studies of Cloud services performance evaluation, however, we found that there was a lack of systematic approaches to factor selection for experimental design. In most cases, evaluators identified factors either randomly or intuitively, and thus prepared evaluation experiments through an ad hoc way. For example, when it comes to the performance evaluation of Amazon EC2, different studies casually considered different EC2 instance factors in different experiments, such as VM type \cite{Stantchev_2009}, number \cite{Stantchev_2009}, geographical location \cite{Iosup_Yigitbasi_2010}, operation system (OS) brand \cite{Li_Yang_2010}, and even CPU architecture \cite{Iosup_Yigitbasi_2010} and brand \cite{Napper_Bientinesi_2009}, etc. In fact, to the best of our knowledge, none of the current Cloud performance evaluation studies has used ``experimental factors" deliberately to design evaluation experiments and analyze the experimental results.

Therefore, we decided to establish a framework of suitable experimental factors to facilitate applying experimental design techniques to the Cloud services evaluation work. Unfortunately, it is difficult to directly point out a full scope of experimental factors for evaluating performance of Cloud services, because the Cloud nowadays is still chaotic compared with traditional computing systems \cite{Stokes_2011}. Consequently, we used a regression manner to construct this factor framework. In other words, we tried to isolate the de facto experimental factors from the state-of-the-practice of Cloud services performance evaluation. In fact, the establishment of this factor framework is a continuation of our previous work \cite{Li_OBrien_2012a,Li_OBrien_2012b,Li_OBrien_2012c} that collected, clarified and rationalized the key concepts and their relationships in the existing Cloud performance evaluation studies. Benefitting from such a de facto factor framework, new evaluators can explore and refer to the existing evaluation concerns for designing their own experiments for performance evaluation of commercial Cloud services.

Note that, as a continuation of our previous work, this study conventionally employed four constrains, as listed below.

\begin{itemize*}
    \item	We focused on the evaluation of only commercial Cloud services, rather than that of private or academic Cloud services, to make our effort closer to industry's needs.
    \item	We only investigated Performance evaluation of commercial Cloud services. The main reason is that not enough data of evaluating the other service features could be found to support the generalization work. For example, there are little empirical studies in Security evaluation of commercial Cloud services due to the lack of quantitative metrics \cite{Li_OBrien_2012b}. 
    \item	We considered Infrastructure as a Service (IaaS) and Platform as a Service (PaaS) without considering Software as a Service (SaaS). Since SaaS with special functionalities is not used to further build individual business applications \cite{Binnig_Kossmann_2009}, the evaluation of various SaaS instances could comprise an infinite and exclusive set of factors that would be out of the scope of this investigation.
    \item	We only explored empirical evaluation practices in academic publications. There is no doubt that informal descriptions of Cloud services evaluation in blogs and technical websites can also provide highly relevant information. However, on the one hand, it is impossible to explore and collect useful data from different study sources all at once. On the other hand, the published evaluation reports can be viewed as typical and peer-reviewed representatives of the existing ad hoc evaluation practices.
\end{itemize*}

The remainder of this paper is organized as follows. Section \ref{II} briefly introduces the four-step methodology that we have used to establish this factor framework. Section \ref{III} specifies the tree-structured factor framework branch by branch. An application case is employed in Section \ref{IV} to demonstrate how the proposed factor framework can help facilitate experimental design for Cloud services performance evaluation. Conclusions and some future work are discussed in Section \ref{V}.

\section{Methodology of Establishing the Framework}
\label{II}
As previously mentioned, this factor framework is established based on our previous work, which is mainly composed of four steps, as listed below and respectively specified in the following subsections:
\begin{itemize*}
    \item	Conduct a systematic literature review (SLR).
    \item	Construct a taxonomy based on the SLR.
    \item	Build a conceptual model based on the taxonomy.
    \item	Establish an experimental factor framework at last.
\end{itemize*}

\subsection{Conduct Systematic Literature Review}
\begin{figure}[!t]
\centering
\includegraphics{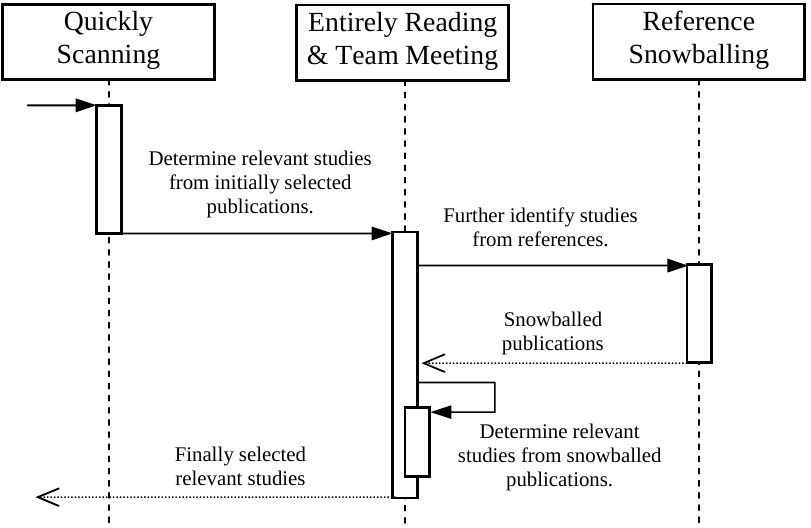}
\caption{The study selection sequence in the SLR on evaluating commercial Cloud services.}
\label{fig>PicSequenceDiagram}
\end{figure}

The foundation for establishing this factor framework is a systematic literature review (SLR) on evaluating commercial Cloud services.\footnote{The complete SLR report can be found online: \url{https://docs.google.com/open?id=0B9KzcoAAmi43LV9IaEgtNnVUenVXSy1FWTJKSzRsdw}} As the main methodology applied for Evidence-Based Software Engineering (EBSE) \cite{Dyba_Kitchenham_2005}, SLR has been widely accepted as a standard and systematic approach to investigation of specific research questions by identifying, assessing, and analyzing published primary studies. Following a rigorous selection process in this SLR, as illustrated in Figure \ref{fig>PicSequenceDiagram}, we have identified 46 Cloud services evaluation studies covering six commercial Cloud providers, such as Amazon, GoGrid, Google, IBM, Microsoft, and Rackspace, from a set of popular digital publication databases (all the identified evaluation studies have been listed online for reference: \url{http://www.mendeley.com/groups/1104801/slr4cloud/papers/}). The evaluation experiments in those identified 46 studies were thoroughly analyzed. In particular, the atomic experimental components, such as evaluation requirements, Cloud service features, metrics, benchmarks, experimental resources, and experimental operations, were respectively extracted and arranged. 

\subsection{Construct Taxonomy}
\begin{figure}[!t]
\centering
\includegraphics{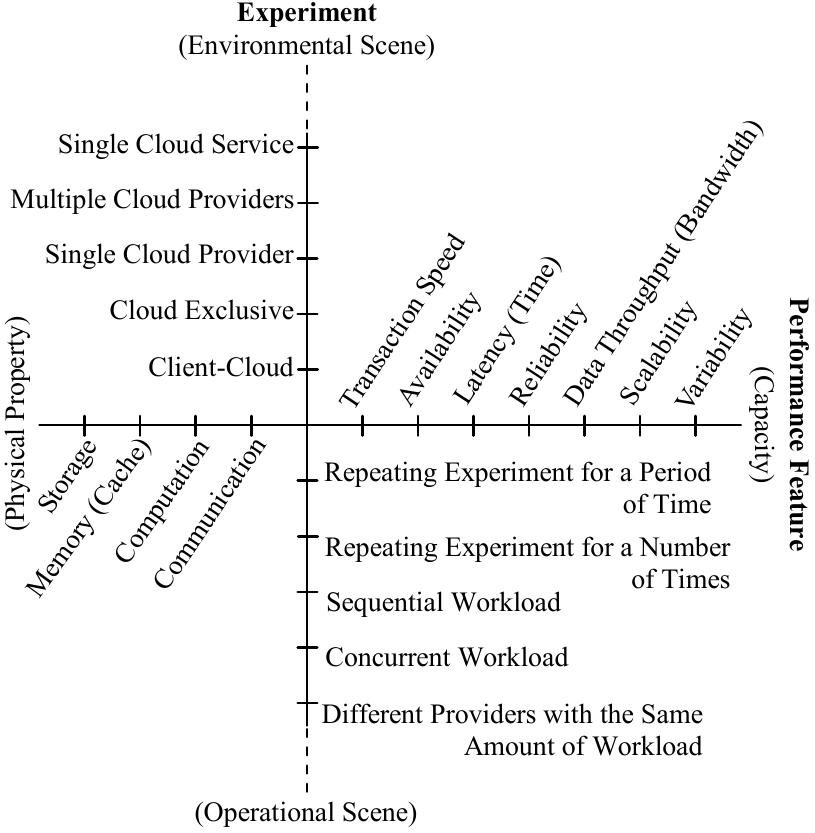}
\caption{Two-dimensional taxonomy of performance evaluation of commercial Cloud services.}
\label{fig>PicTaxonomy}
\end{figure}

During the analysis of these identified evaluation studies, we found that there were frequent reporting issues ranging from non-standardized specifications to misleading explanations \cite{Li_OBrien_2012a}. Considering that those issues would inevitably obstruct comprehending and spoil drawing lessons from the existing evaluation work, we created a novel taxonomy to clarify and arrange the key concepts and terminology for Cloud services performance evaluation. The taxonomy is constructed along two dimensions: Performance Feature and Experiment. Moreover, the Performance Feature dimension is further split into \textit{Physical Property} and \textit{Capacity} parts, while the Experiment dimension is split into \textit{Environmental Scene} and \textit{Operational Scene} parts, as shown in Figure \ref{fig>PicTaxonomy}. The details of this taxonomy has been elaborated in \cite{Li_OBrien_2012a}.

\subsection{Build Conceptual Model}
\label{II>model}
\begin{figure}[!t]
\centering
\includegraphics{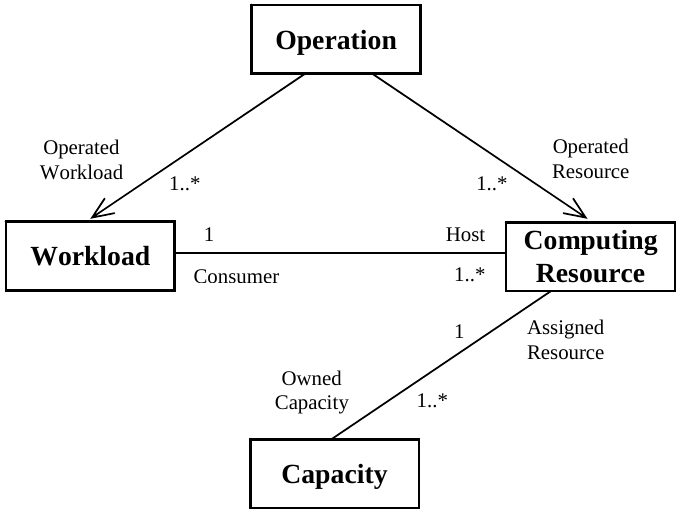}
\caption{Conceptual model of Cloud services performance evaluation in the top classifier layer.}
\label{fig>PicEvaluationModel}
\end{figure}

Since a model is an abstract summary of some concrete object or activity in reality \cite{Mellor_Clark_2003}, the identification of real and concrete objects/activities plays a fundamental role in the corresponding modeling work. Given that the taxonomy had capsuled relevant key concepts and terminology, we further built a conceptual model of performance evaluation of commercial Cloud services to rationalize different abstract-level classifiers and their relationships \cite{Li_OBrien_2012c}. In detail, we used a three-layer structure to host different abstract elements for the performance evaluation conceptual model. To save space, here we only portray the most generalized part hosted in the top classifier layer, as shown in Figure \ref{fig>PicEvaluationModel}, which reflects the most generic reality of performance evaluation of a computing paradigm: essentially, performance evaluation can be considered as \emph{exploring the capacity of particular computing resources with particular workloads driven by a set of operations}.

\subsection{Establish Factor Framework}
In fact, the specific classifiers in the abovementioned conceptual model \cite{Li_OBrien_2012c} has implied the state-of-the-practice of performance evaluation factors that people currently took into account in the Cloud Computing domain. According to different positions in the process of an evaluation experiment \cite{Antony_2003}, the specific classifiers of Workload and Computing Resource indicate input process factors; the specific classifiers of Capacity suggest output process factors; while the Operation classifiers are used to adjust values of input process factors. The detailed experimental factors for Cloud services performance evaluation are elaborated in the next section.\\

\section{The Tree-structured Factor Framework}
\label{III}
As mentioned previously, the experimental factors for performance evaluation of commercial Cloud services can be categorized into two input process groups (Workload and Computing Resource) and one output process group (Capacity). Thus, we naturally portrayed the factor framework as a tree with three branches. Each of the following subsections describes one branch of the factor tree.
\subsection{Workload Factors}
\begin{figure}[!t]
\centering
\includegraphics{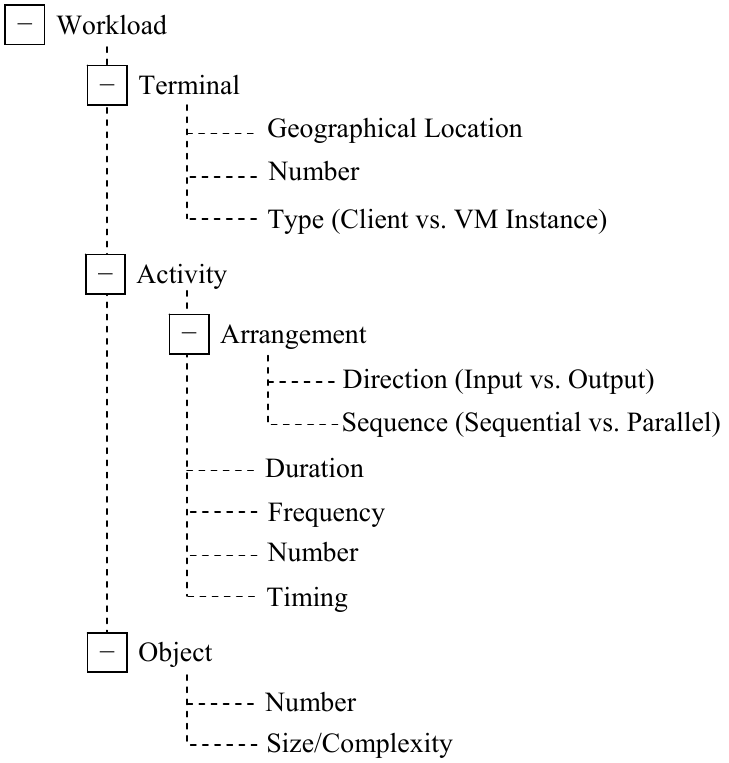}
\caption{The workload factors for experimental design.}
\label{fig>PicWorkloadTree}
\end{figure}

Based on our previous work \cite{Li_OBrien_2012a,Li_OBrien_2012c}, we found that a piece of workload used in performance evaluation could be described through one of three different concerns or a combination of them, namely Terminal, Activity, and Object. As such, we can adjust the workload by varying any of the concerns through different experimental operations. The individual workload factors are listed in Figure \ref{fig>PicWorkloadTree}.

\subsubsection{Terminal}
In contrast with services to be evaluated in the Cloud, clients and particular Cloud resource (usually VM instances) issuing workload activities can be viewed as terminals. Correspondingly, the \textit{geographical location} or \textit{number} of both clients \cite{Garfinker_2007} and VM instances \cite{Hill_Li_2010} have been used to depict the relevant workload. Meanwhile, the \textit{terminal type} can also be used as a workload factor. For example, the authors evaluated Cloud network latency by using client and EC2 instance respectively to issue pings \cite{Baun_Kunze_2009}. In this case, the \textit{terminal type} has the equal essence to the factor \textit{communication scope} (cf.~Subsection \ref{III>Commun}).  
\subsubsection{Activity}
The concept ``activity" here describes an inherent property of workload, which is different from, but adjustable by, experimental operations. For example, disk I/O request as a type of activity can be adjusted by operations like the number or time of the requests. In fact, the number- and time-related variables, such as \textit{activity duration} \cite{Garfinker_2007}, \textit{frequency} \cite{Chiu_Agrawal_2010}, \textit{number} \cite{Chiu_Agrawal_2010}, and \textit{timing} \cite{Garfinker_2007a}, have been widely considered as workload factors in practice. Furthermore, by taking a particular Cloud resource being evaluated as a reference, the factor \textit{activity direction} can be depicted as input or output \cite{Baun_Kunze_2009}. As for the \textit{activity sequence} in a workload, the arrangement generates either sequential \cite{Baun_Kunze_2009} or parallel \cite{Hill_Li_2010} activity flows.
\subsubsection{Object}
In a workload for Cloud services performance evaluation, objects refer to the targets of the abovementioned activities. The concrete objects can be individual messages \cite{Hill_Humphrey_2009}, data files \cite{Hill_Li_2010}, and transactional jobs/tasks \cite{Deelman_Singh_2008} in fine grain, while they can also be coarse-grained workflows or problems \cite{Deelman_Singh_2008}. Therefore, the \textit{object number} and \textit{object size/complexity} are two typical workload factors in the existing evaluation studies. Note that we do not consider object location as a workload factor, because the locations of objects are usually hosted and determined by computing resources (cf.~Subsection \ref{III>resources}). In particular, a workload may have multiple object size/complexity-related factors in one experiment. For example, a set of parameters of HPL benchmark, such as the block size and process grid size, should be tuned simultaneously when evaluating Amazon EC2 \cite{Bientinesi_Iakymchuk_2010}.

\subsection{Computing Resource Factors}
\label{III>resources}
\begin{figure}[!t]
\centering
\includegraphics{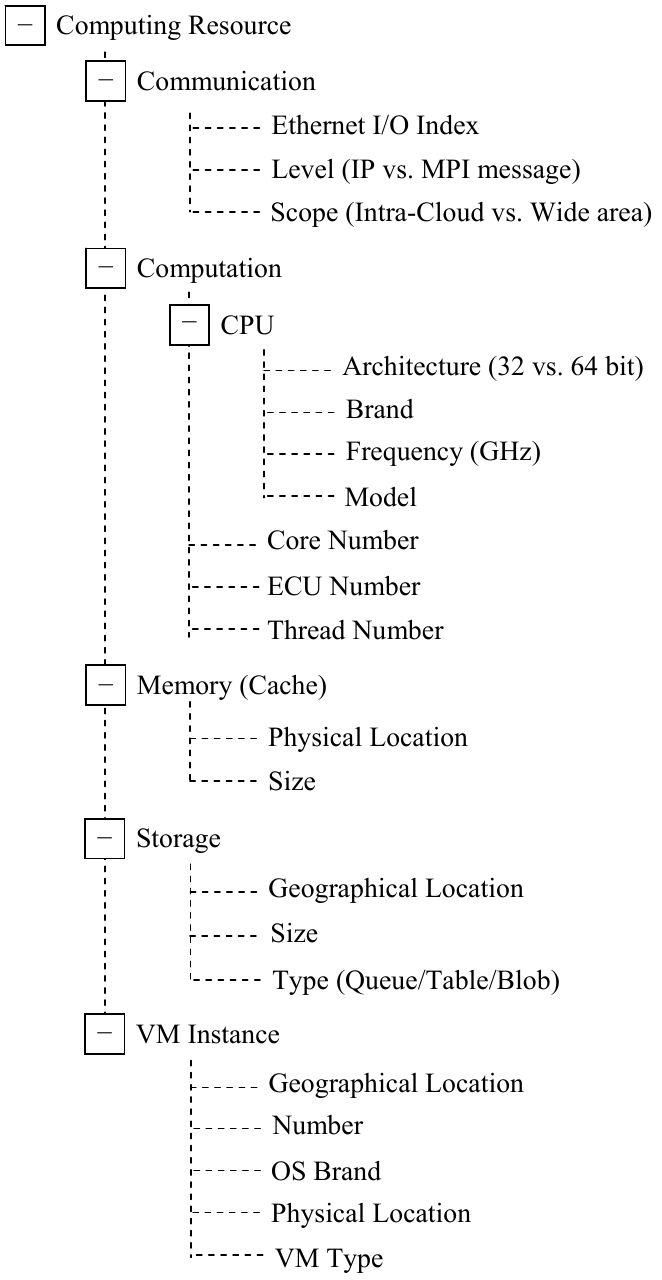}
\caption{The computing resource factors for experimental design.}
\label{fig>PicResourceTree}
\end{figure}

According to the physical properties in the performance feature of commercial Cloud services \cite{Li_OBrien_2012a}, the Cloud Computing resource can be consumed by one or more of four basic styles: Communication, Computation, Memory (Cache), and Storage. In particular, the VM Instance resource is an integration of all the four basic types of computing resources. Overall, the computing resource factors can be organized as Figure \ref{fig>PicResourceTree} shows.

\subsubsection{Communication}
\label{III>Commun}
As explained in \cite{Li_OBrien_2012a}, Communication becomes a special Cloud Computing resource because commercial Cloud services are employed inevitably through Internet/Ethernet. As such, the \textit{Ethernet I/O Index} is usually pre-supplied as a service-level agreement (SLA) by service providers. In practice, the scope and level of communication have been frequently emphasized in the performance evaluation studies. Therefore, we can summarize two practical factors: The factor \textit{Communication Scope} considers intra-Cloud and wide-area data transferring respectively \cite{Li_Yang_2010}, while the \textit{Communication Level} distinguishes between IP-level and MPI-message-level networking \cite{He_Zhou_2010}.

\subsubsection{Computation}
When evaluating PaaS, the Computation resource is usually regarded as a black box \cite{Iosup_Yigitbasi_2010}. Whereas, for IaaS, the practices of Computation evaluation of Cloud services have taken into account \textit{Core Number} \cite{Bientinesi_Iakymchuk_2010}, \textit{Elastic Compute Unit (ECU) Number}, \textit{Thread Number} \cite{Baun_Kunze_2009}, and a set of CPU characteristics. Note that, compared to physical CPU core and thread, ECU is a logical concept introduced by Amazon, which is defined as the CPU power of a 1.0-1.2 GHz 2007 Opteron or Xeon processor \cite{Ostermann_Iosup_2009}. When it comes to CPU characteristics, the \textit{Architecture} (e.g. 32 bit vs. 64 bit \cite{Iosup_Yigitbasi_2010}) and \textit{Brand} (e.g. AMD Opteron vs. Intel Xeon \cite{Napper_Bientinesi_2009}) have been respectively considered in evaluation experiments. Processors with the same brand can be further distinguished between different \textit{CPU Models} (e.g. Intel Xeon E5430 vs. Intel Xeon X5550 \cite{Bientinesi_Iakymchuk_2010}). In particular, \textit{CPU Frequency} appears also as an SLA of Cloud computation resources.

\subsubsection{Memory (Cache)}
Since Memory/Cache could closely work with the Computation and Storage resources in computing jobs, it is hard to exactly distinguish the affect to performance brought by Memory/Cache. Therefore, not many dedicated Cloud memory/cache evaluation studies can be found from the literature. In addition to the SLA \textit{Memory Size}, interestingly, \textit{Physical Location} and \textit{Size} of cache (e.g. L1=64KB vs. L2=1MB in Amazon m1.* instances \cite{Ostermann_Iosup_2009}) have attracted attentions when analyzing the memory hierarchy. However, in \cite{Ostermann_Iosup_2009}, different values of these factors were actually revealed by performance evaluation rather than used for experimental design.

\subsubsection{Storage}
As mentioned in \cite{Li_OBrien_2012a}, Storage can be either the only functionality or a component functionality of a Cloud service, for example Amazon S3 vs. EC2. Therefore, it can be often seen that disk-related storage evaluation also adopted experimental factors of evaluating other relevant resources like VM instances (cf.~Subsection \ref{III-VM}). Similarly, the predefined \textit{Storage Size} acts as an SLA, while a dedicated factor of evaluating Storage is the \textit{Geographical Location}. Different geographical locations of Storage resources can result either from different service data centers (e.g. S3 vs. S3-Europe \cite{Palankar_Iamnitchi_2008}) or from different storing mechanisms (e.g. local disk vs. remote NFS drive \cite{Sobel_Subramanyam_2008}). In addition, although not all of the public Cloud providers specified the definitions, the Storage resource has been distinguished among three types of offers: Blob, Table and Queue \cite{Li_Yang_2010}. Note that different \textit{Storage Types} correspond to different sets of data-access activities, as described in \cite{Li_OBrien_2012b}. 

\subsubsection{VM Instance}
\label{III-VM}
VM Instance is one of the most popular computing resource styles in the commercial Cloud service market. The widely considered factors in current VM Instance evaluation experiments are \textit{Geographical Location}, \textit{Instance Number}, and \textit{VM Type} \cite{Bientinesi_Iakymchuk_2010,Hill_Humphrey_2009,Hill_Li_2010,Iosup_Yigitbasi_2010,Li_Yang_2010,Ostermann_Iosup_2009,Stantchev_2009}. The \textit{VM Type} of a particular instance naturally reflects its corresponding provider, as demonstrated in \cite{Li_Yang_2010}. Moreover, although not common, the \textit{OS Brand} (e.g. Linux vs. Windows \cite{Li_Yang_2010}) and \textit{Physical Location} \cite{Dejun_Pierre_2009} also emerged as experimental factors in some evaluation studies. Note that the physical location of a VM instance indicates the instance's un-virtualized environment, which is not controllable by evaluators in evaluation experiments \cite{Dejun_Pierre_2009}. In particular, recall that a VM Instance integrates above four basic types of computing resources. We can therefore find that some factors of evaluating previous resources were also used in the evaluation of VM Instances, for example the \textit{CPU Architecture} and \textit{Core Number} \cite{Bientinesi_Iakymchuk_2010,Ostermann_Iosup_2009}.

\subsection{Capacity Factors}
\begin{figure}[!t]
\centering
\includegraphics{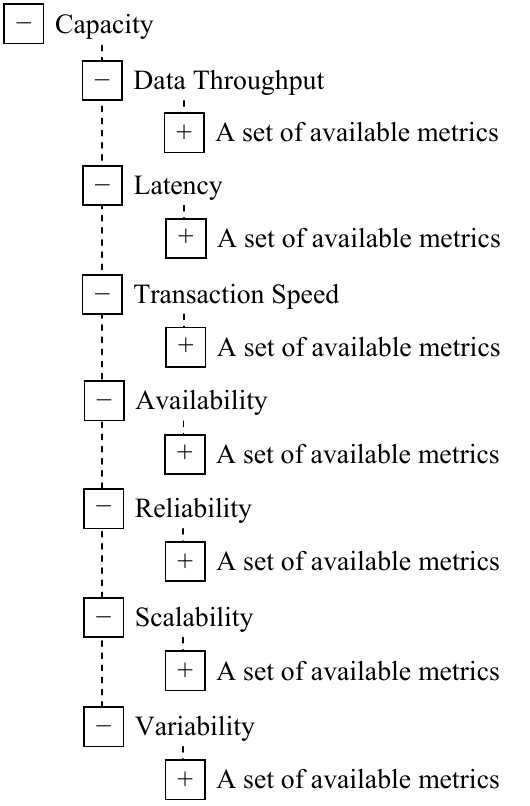}
\caption{The capacity factors for experimental design.}
\label{fig>PicCapacityTree}
\end{figure}

As discussed about the generic reality of performance evaluation in Subsection \ref{II>model}, it is clear that the capacities of a Cloud computing resource are intangible until they are measured. Meanwhile, the measurement has to be realized by using measurable and quantitative metrics \cite{Le_Boudec_2011}. Therefore, we can treat the values of relevant metrics as tangible representations of the evaluated capacities. Moreover, a particular capacity of a commercial Cloud service may be reflected by a set of relevant metrics, and each metric provides a different lens into the capacity as a whole \cite{Fortier_Michel_2003}. For example, Benchmark Transactional Job Delay \cite{Luckow_Jha_2010} and Benchmark Delay \cite{Juve_Deelman_2009} are both Latency metrics: the former is from the individual perspective, while the latter from the global perspective.
As such, we further regard relevant metrics as possible output process factors \cite{Antony_2003} when measuring a particular Cloud service capacity, and every single output process factor can be used as a candidate response \cite{Antony_2003} in the experimental design. Since we have clarified seven different Cloud service capacities \cite{Li_OBrien_2012a}, i.e.~Data Throughput, Latency, Transaction Speed, Availability, Reliability, Scalability, and Variability, the possible capacity factors (metrics) can be correspondingly categorized as Figure \ref{fig>PicCapacityTree} shows. Due to the limit of space, it is impossible and unnecessary to exhaustively list all the metrics in this paper. In fact, the de facto metrics for performance evaluation of commercial Cloud services have been collected and summarized in our previous work \cite{Li_OBrien_2012b}.

\section{Application of the Factor Framework}
\label{IV}
Since the factor framework is inherited from the aforementioned taxonomy and modeling work, it can also be used for, and in turn be validated by, analyzing the existing studies of Cloud services performance evaluation, as described in \cite{Li_OBrien_2012a,Li_OBrien_2012c}. To avoid duplication, we do not elaborate the analysis application scenario, and the corresponding validation, of the factor framework in this paper. Instead, we particularly highlight and demonstrate how this factor framework can help facilitate designing experiments for evaluating performance of commercial Cloud services.

Suppose there is a requirement of evaluating Amazon EC2 with respect to its disk I/O. Recall that relevant factors play a prerequisite role in designing evaluation experiments. Given the factor framework proposed in this paper, we can quickly and conveniently lookup and choose experimental factors according to the evaluation requirement. To simplify the demonstration, here we constrain the terminal to be clients, while only consider the direction of disk I/O and data size to be read/write in workload factors, and only consider the EC2 VM type in computing resource factors. As for the capacity factors, we can employ multiple suitable metrics in this evaluation, for example disk I/O latency and data throughput. However, since only one metric should be determined as the response in an experimental design \cite{Antony_2003}, we choose the disk data throughput in this case. Thus, we have circled \textit{active direction}, \textit{object size}, and \textit{VM type} as factors, while \textit{data throughput} as response in the framework for designing experiments. In particular, we use two-level settings for the three factors: the value of \textit{active direction} can be Write or Read; \textit{object size} can be Char or Block; and \textit{VM type} only covers M1.small and M1.large. In addition, we use ``MB/s" as the unit of \textit{data throughput}.

Since only a small amount of factors are concerned, we can simply adopt the most straightforward design technique, namely Full-factorial Design \cite{Antony_2003}, for this demonstration. This design technique adjusts one factor at a time, which results in an experimental matrix comprising eight trials, as shown in Matrix (\ref{eq>1}). For conciseness, we further assign aliases to those experimental factors, as listed below. Note that the sequence of the experimental trials has been randomized to reduce possible noises or biases \cite{Antony_2003} during the designing process. 

\begin{itemize*}
    \item	A: Activity Direction (Write vs. Read).
    \item	B: Object Size (Char vs. Block). 
    \item	C: VM Type (M1.small vs. M1.large).
    \item	Response: Data Throughput (MB/s).
\end{itemize*}

\begin{small}
\begin{equation}
\label{eq>1}
\left[
  \begin{array}{ccccc}
    trial & A & B & C & Response\\
    1 & Write & Block & M1.small & ? \\
    2 & Read & Char & M1.large & ? \\
    3 & Write & Char & M1.small & ?\\
    4 & Read & Char & M1.small & ?\\
    5 & Read & Block & M1.large & ?\\
    6 & Read & Block & M1.small & ?\\
    7 & Write & Char & M1.large & ?\\
    8 & Write & Block & M1.large & ?\\
  \end{array}
\right]
\end{equation} 
\end{small}

Following the experimental matrix, we can implement evaluation experiments trial by trial, and fill the Response column with experimental results. For our convenience, here we directly employ the evaluation results reported in \cite{Iosup_Ostermann_2011}, as listed in Matrix (\ref{eq>2}).

\begin{small}
\begin{equation}
\label{eq>2}
\left[
  \begin{array}{ccccc}
    trial & A & B & C & Response\\
    1 & Write & Block & M1.small & 73.5~MB/s \\
    2 & Read & Char & M1.large & 50.9~MB/s \\
    3 & Write & Char & M1.small & 25.9~MB/s\\
    4 & Read & Char & M1.small & 22.3~MB/s\\
    5 & Read & Block & M1.large & 64.3~MB/s\\
    6 & Read & Block & M1.small & 60.2~MB/s\\
    7 & Write & Char & M1.large & 35.9~MB/s\\
    8 & Write & Block & M1.large & 63.2~MB/s\\
  \end{array}
\right]
\end{equation} 
\end{small}

\begin{figure}[!t]
\centering
\includegraphics[width=3in]{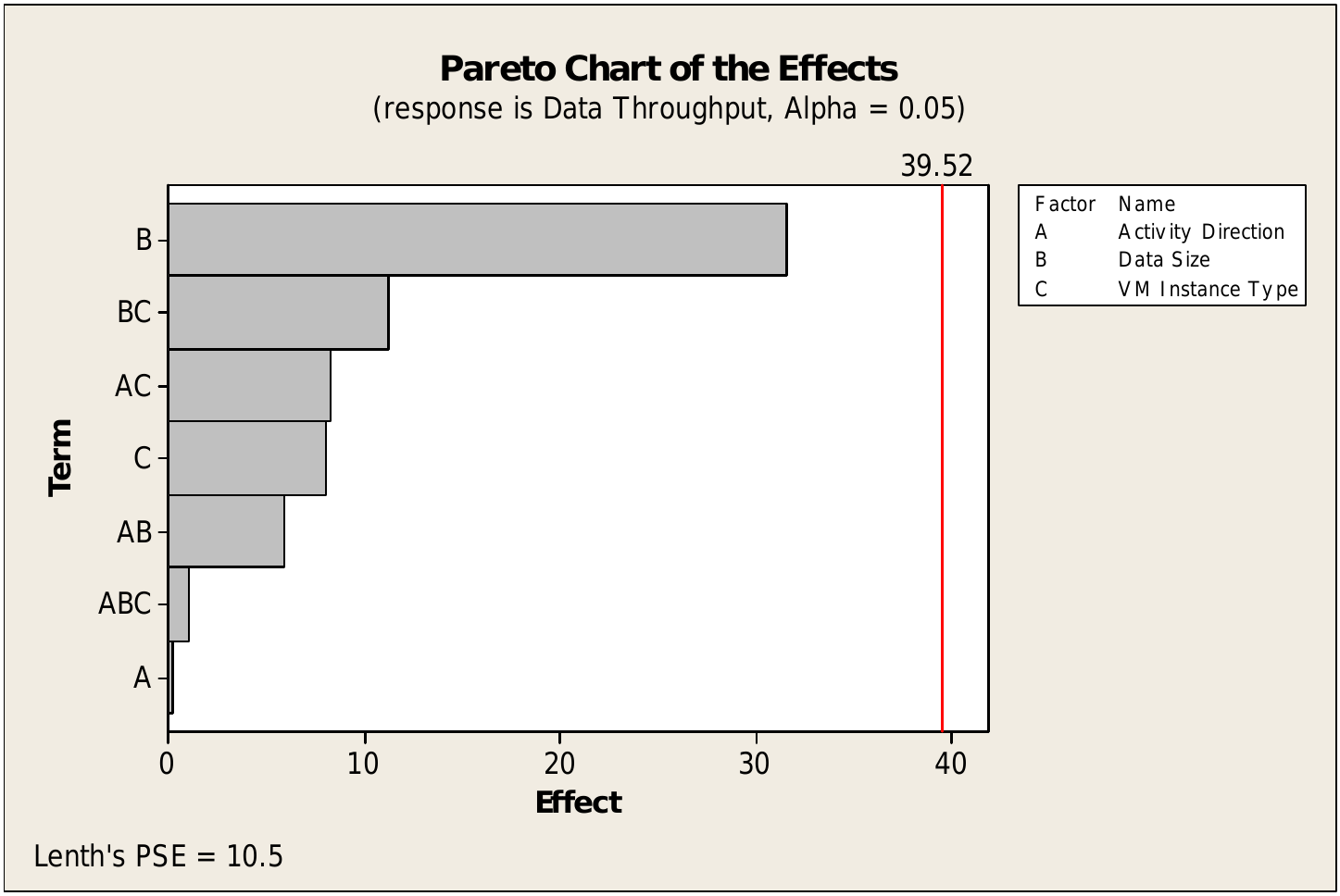}
\caption{The Pareto plot of factor effects.}
\label{fig>PicPlot}
\end{figure}

Finally, different analytical techniques can be employed to reveal more comprehensive meanings of experimental results \cite{Antony_2003} for commercial Cloud services. For example, in this case, we can further investigate the significances of these factors to analyze their different influences on the disk I/O performance. In detail, by setting the significance level $\alpha$ as 0.05 \cite{Jackson_2011}, we draw a Pareto plot to detect the factor and interaction effects that are important to the process of reading/writing data from/to EC2 disks, as shown in Figure \ref{fig>PicPlot}. 

Given a particular significance level, Pareto plot displays a red reference line besides the effect values. Any effect that extends past the reference line is potentially important \cite{Antony_2003}. In Figure \ref{fig>PicPlot}, none of the factor or interaction effects is beyond the reference line, which implies that none of the factors or interactions significantly influences the EC2 disk I/O performance. Therefore, we can claim that EC2 disk I/O is statistically stable with respect to those three factors. However, Factor B (Data Size to be read/written) has relatively significant influence on the performance of EC2 disk I/O. Since the throughput of small-size data (Char) is much lower than that of large-size data (Block), we can conclude that there is a bottleneck of transaction overhead when reading/writing small size of data. On the contrary, there is little I/O performance effect when switching activity directions, which means the disk I/O of EC2 is particularly stable no matter reading or writing the same size of data.

Overall, through the demonstration, we can find that this factor framework offers a concrete and rational foundation for implementing performance evaluation of commercial Cloud services. When evaluating Cloud services, there is no doubt that the techniques of experimental design and analysis can still be applied by using intuitively selected factors. Nevertheless, by referring to the existing evaluation experiences, evaluators can conveniently identify suitable experimental factors while excluding the others, which essentially suggests a systematic rather than ad hoc decision making process.

\section{Conclusions and Future Work}
\label{V}
Cloud Computing has attracted tremendous amount of attention from both customers and providers in the current computing industry, which leads to a competitive market of commercial Cloud services. As a result, different Cloud infrastructures and services may be offered with different terminology, definitions, and goals \cite{Prodan_Ostermann_2009}. On one hand, different Cloud providers have their own idiosyncratic characteristics when developing services \cite{Li_Yang_2010}. On the other hand, even the same provider can supply different Cloud services with comparable functionalities for different purposes. For example, Amazon has provided several options of storage service, such as EC2, EBS, and S3 \cite{Chiu_Agrawal_2010}. Consequently, performance evaluation of candidate services would be crucial and beneficial for many purposes ranging from cost-benefit analysis to service improvement \cite{Li_Yang_2010}.

When it comes to performance evaluation of a computing system, proper experiments should be designed with respect to a set of factors that may influence the system's performance \cite{Jain_1991}. In the Cloud Computing domain, however, we could not find any performance evaluation study intentionally concerning ``factors" for experimental design and analysis. On the contrary, most of the evaluators intuitively employed experimental factors and prepared ad hoc experiments for evaluating performance of commercial Cloud services. Considering factor identification plays a prerequisite role in experimental design, it is worthwhile and necessary to investigate the territory of experimental factors to facilitate evaluating Cloud services more systematically. Therefore, based on our previous work, we collected experimental factors that people currently took into account in Cloud services performance evaluation, and arrange them into a tree-structured framework.

The most significant contribution of this work is that the framework supplies a dictionary-like approach to selecting experimental factors for Cloud services performance evaluation. Benefitting from the framework, evaluators can identify necessary factors in a concrete space instead of on the fly. In detail, as demonstrated in the EC2 disk I/O evaluation case in Section \ref{IV}, given a particular evaluation requirement, we can quickly and conveniently lookup and circle relevant factors in the proposed framework to design evaluation experiments, and further analyze the effects of the factors and their interactions to reveal more of the essential nature of the evaluated service. Note that this factor framework is supposed to supplement, but not replace, the expert judgement for experimental factor identification, which would be particularly helpful for Cloud services evaluation when there is a lack of a bunch of experts.


The future work of this research will be unfolded along two directions. First, we will gradually collect feedback from external experts to supplement this factor framework. As explained previously, Cloud Computing is still maturing and relatively chaotic \cite{Stokes_2011}, it is therefore impossible to exhaustively identify the relevant experimental factors all at once. Through smooth expansion, we can make this factor framework increasingly suit the more general area of evaluation of Cloud Computing. Second, given the currently available factors, we plan to formally introduce and adapt suitable techniques of experimental design and analysis to evaluating commercial Cloud services. With experimental design and analysis techniques, this factor framework essentially acts as a solid base to support systematic implementations of Cloud services evaluation. 

\section*{Acknowledgment}

This project is supported by the Commonwealth of Australia under the Australia-China Science and Research Fund.

NICTA is funded by the Australian Government as represented by the Department of Broadband, Communications and the Digital Economy and the Australian Research Council through the ICT Centre of Excellence program.




\begin{thebibliography}{99}
\itemsep 3pt

\bibitem{Antony_2003}
J.~Antony, \emph{Design of Experiments for Engineers and Scientists}. Burlington, MA: Butterworth-Heinemann, Nov.~2003.

\bibitem{Baun_Kunze_2009}
C.~Baun and M.~Kunze, ``Performance Measurement of a Private Cloud in the OpenCirrus\textsuperscript{TM} Testbed," \emph{Proc.~4th Workshop on Virtualization in High-Performance Cloud Computing (VHPC 2009) in conjunction with 15th Int.~European Conf.~Parallel and Distributed Computing (Euro-Par 2009)}, Springer-Verlag, Aug.~2009, pp.~434--443.

\bibitem{Bientinesi_Iakymchuk_2010}
P.~Bientinesi, R.~Iakymchuk, and J.~Napper, ``HPC on competitive Cloud resources," \emph{Handbook of Cloud Computing}, B.~Furht and A.~Escalante, eds., New York: Springer-Verlag, pp.~493--516, 2010.

\bibitem{Binnig_Kossmann_2009}
C.~Binnig, D.~Kossmann, T.~Kraska, and S.~Loesing, ``How is the weather tomorrow? Towards a benchmark for the Cloud," \emph{Proc.~2nd Int.~Workshop on Testing Database Systems (DBTest 2009) in conjunction with ACM SIGMOD/PODPS Int.~Conf.~Management of Data (SIGMOD/PODS 2009)}, ACM Press, Jun.~2009, pp.~1--6.

\bibitem{Chiu_Agrawal_2010}
D.~Chiu and G.~Agrawal, ``Evaluating caching and storage options on the Amazon Web services Cloud," \emph{Proc.~11th ACM/IEEE Int.~Conf.~Grid Computing (Grid 2010)}, IEEE Computer Society, Oct.~2010, pp.~17--24.

\bibitem{Deelman_Singh_2008}
E.~Deelman, G.~Singh, M.~Livny, B.~Berriman, and J.~Good, ``The cost of doing science on the Cloud: The montage example," \emph{Proc.~2008 Int.~Conf.~High Performance Computing, Networking, Storage and Analysis (SC 2008)}, IEEE Computer Society, Nov.~2008, pp.~1--12.

\bibitem{Dejun_Pierre_2009}
J.~Dejun, G.~Pierre, and C.-H.~Chi, ``EC2 Performance Analysis for Resource Provisioning of Service-Oriented Applications," \emph{Proc.~2009 Int.~Conf.~Service-Oriented Computing (ICSOC/ServiceWave 2009)}, Springer-Verlag, Nov.~2009, pp.~197--207.

\bibitem{Dyba_Kitchenham_2005}
T. Dyb\aa, B. A. Kitchenham, and M. J\o rgensen, ``Evidence-based software engineering for practitioners," \emph{IEEE Softw.}, vol.~22, no.~1, Jan.~2005, pp.~58--65.

\bibitem{Fortier_Michel_2003}
P.J.~Fortier and H.E.~Michel, \emph{Computer Systems Performance Evaluation and Prediction}. Burlington, MA: Digital Press, Jul.~2003.

\bibitem{Garfinker_2007a}
S.~Garfinkel, ``Commodity grid computing with Amazon's S3 and EC2," \emph{;Login}, vol. 32, no. 1, Feb.~2007, pp.~7--13.

\bibitem{Garfinker_2007}
S.~Garfinkel, ``An evaluation of Amazon's grid computing services: EC2, S3 and SQS," Sch.~Eng.~Appl.~Sci., Harvard Univ., Canbridge, MA, Tech.~Rep.~TR-08-07, Jul.~2007.

\bibitem{He_Zhou_2010}
Q.~He, S.~Zhou, B.~Kobler, D.~Duffy, and T.~McGlynn, ``Case study for running HPC applications in public Clouds," \emph{Proc.~19th ACM Int.~Symp.~High Performance Distributed Computing (HPDC 2010)}, ACM Press, Jun.~2010, pp.~395--401.

\bibitem{Hill_Humphrey_2009}
Z.~Hill and M.~Humphrey, ``A quantitative analysis of high performance computing with Amazon's EC2 infrastructure: The death of the local cluster?," \emph{Proc.~10th IEEE/ACM Int.~Conf.~Grid Computing (Grid 2009)}, IEEE Computer Society, Oct.~2009, pp.~26--33.

\bibitem{Hill_Li_2010}
Z.~Hill, J.~Li, M.~Mao, A.~Ruiz-Alvarez, and M.~Humphrey, ``Early observations on the performance of Windows Azure," \emph{Proc.~19th ACM Int.~Symp.~High Performance Distributed Computing (HPDC 2010)}, ACM Press, Jun.~2010, pp.~367--376.

\bibitem{Iosup_Ostermann_2011}
A.~Iosup, S.~Ostermann, N.~Yigitbasi, R.~Prodan, T.~Fahringer, and D.H.J.~Epema, ``Performance analysis of Cloud computing services for many-tasks scientific computing," \emph{IEEE Trans.~Parallel Distrib.~Syst.}, vol.~22, no.~6, Jun.~2011, pp.~931--945.

\bibitem{Iosup_Yigitbasi_2010}
A.~Iosup, N.~Yigitbasi, and D.~Epema, ``On the performance variability of production Cloud services," Delft Univ.~Technol., Netherlands, Tech.~Rep.~PDS-2010-002, Jan.~2010.

\bibitem{Jackson_2011}
S.L.~Jackson, \emph{Research Methods and Statistics: A Critical Thinking Approach}, 4th ed. Belmont, CA: Wadsworth Publishing, Mar.~2011.

\bibitem{Jain_1991}
R.K.~Jain, \emph{The Art of Computer Systems Performance Analysis: Techniques for Experimental Design, Measurement, Simulation, and Modeling}. New York, NY: Wiley Computer Publishing, John Wiley \& Sons, Inc., May 1991.

\bibitem{Juve_Deelman_2009}
G.~Juve, E.~Deelman, K.~Vahi, G.~Mehta, B.~Berriman, B.P.~Berman, and P.~Maechling, ``Scientific workflow applications on Amazon EC2," \emph{Proc.~Workshop on Cloud-based Services and Applications in conjunction with the 5th IEEE Int.~Conf. e-Science (e-Science 2009)}, IEEE Computer Society, Dec.~2009, pp.~59--66.

\bibitem{Le_Boudec_2011}
J.-Y.~Le Boudec, \emph{Performance Evaluation of Computer and Communication Systems}. Lausanne, Switzerland: EFPL Press, Feb.~2011.

\bibitem{Li_OBrien_2012a}
Z.~Li, L.~O'Brien, R.~Cai, and H.~Zhang, ``Towards a taxonomy of performance evaluation of commercial Cloud services," \emph{Proc.~5th Int.~Conf.~Cloud Computing (IEEE CLOUD 2012)}, IEEE Computer Society, Jun.~1012, pp.~344--351.

\bibitem{Li_OBrien_2012b}
Z.~Li, L.~O'Brien, H.~Zhang, and R.~Cai, ``On a catalogue of metrics for evaluating commercial Cloud services," \emph{Proc.~13th ACM/IEEE Int.~Conf.~Grid Computing (Grid 2012)}, IEEE Computer Society, Sept.~2012, pp.~164--173.

\bibitem{Li_OBrien_2012c}
Z.~Li, L.~O'Brien, H.~Zhang, and R.~Cai, ``On a taxonomy-based conceptual model of performance evaluation of Infrastructure as a Service," submitted to \emph{IEEE Trans.~Serv.~Comput.}.

\bibitem{Li_Yang_2010}
A.~Li, X.~Yang, S.~Kandula, and M.~Zhang, ``CloudCmp: Comparing public Cloud providers," \emph{Proc.~10th Annu.~Conf.~Internet Measurement (IMC 2010)}, ACM Press, Nov.~2010 pp.~1--14.

\bibitem{Luckow_Jha_2010}
A.~Luckow and S.~Jha, ``Abstractions for loosely-coupled and ensemble-based simulations on Azure," \emph{Proc.~2nd IEEE Int.~Conf.~Cloud Computing Technology and Science (CloudCom 2010)}, IEEE Computer Society, Nov./Dec.~2010, pp.~550--556.

\bibitem{Mellor_Clark_2003}
S.J.~Mellor, A.N.~Clark, and T.~Futagami, ``Model-driven development - guest editor's introduction," \emph{IEEE Softw.}, vol.~20, no.~5, Sept./Oct.~2003, pp.~14--18.

\bibitem{Napper_Bientinesi_2009}
J.~Napper and P.~Bientinesi, ``Can Cloud computing reach the top500?," \emph{Proc.~Combined Workshops on UnConventional High Performance Computing Workshop plus Memory Access Workshop (UCHPC-MAW 2009)}, ACM Press, May 2009, pp.~17--20.

\bibitem{Ostermann_Iosup_2009}
S.~Ostermann, A.~Iosup, N.~Yigitbasi, R.~Prodan, T.~Fahringer, and D.H.J.~Epema, ``A performance analysis of EC2 Cloud computing services for scientific computing," \emph{Proc.~1st Int.~Conf.~Cloud Computing (CloudComp 2009)}, Springer-Verlag, Oct.~2009, pp.~115--131.

\bibitem{Palankar_Iamnitchi_2008}
M.R.~Palankar, A.~Iamnitchi, M.~Ripeanu, and S.~Garfinkel, ``Amazon S3 for science grids: A viable solution?," \emph{Proc.~2008 Int.~Workshop on Data-aware Distributed Computing (DADC 2008)}, ACM Press, Jun.~2008, pp.~55--64.

\bibitem{Prodan_Ostermann_2009}
R.~Prodan and S.~Ostermann, ``A survey and taxonomy of Infrastructure as a Service and Web hosting Cloud providers," \emph{Proc.~10th IEEE/ACM Int.~Conf.~Grid Computing (Grid 2009)}, IEEE Computer Society, Oct.~2009, pp.~17--25.

\bibitem{Sobel_Subramanyam_2008}
W.~Sobel, S.~Subramanyam, A.~Sucharitakul, J.~Nguyen, H.~Wong, A.~Klepchukov, S.~Patil, A.~Fox, and D.~Patterson, ``Cloudstone: Multi-platform, multi-language benchmark and measurement tools for Web 2.0," \emph{Proc.~1st Workshop on Cloud Computing and Its Applications (CCA 2008)}, Oct.~2008, pp.~1--6.

\bibitem{Stantchev_2009}
V.~Stantchev, ``Performance evaluation of Cloud computing offerings," \emph{Proc.~3rd Int.~Conf.~Advanced Engineering Computing and Applications in Sciences (ADVCOMP 2009)}, IEEE Computer Society, Oct.~2009, pp.~187--192.

\bibitem{Stokes_2011}
J.~Stokes, ``The PC is order, the Cloud is chaos," \emph{Wired Cloudline, Wired.com}, available at \url{http://www.wired.com/cloudline/2011/12/the-pc-is-order/}, Dec.~2011.

\end{thebibliography}
%

\end{document}